\documentclass[%
 aip,
 amsmath,amssymb,
 reprint,%
]{revtex4-1}

\usepackage{graphicx}
\usepackage{dcolumn}
\usepackage{bm}

\usepackage[utf8]{inputenc}
\usepackage[T1]{fontenc}
\usepackage{mathptmx}
\usepackage{etoolbox}

\usepackage[english]{babel}

\usepackage{amsmath}
\usepackage{braket}
\usepackage[colorlinks=true, allcolors=blue]{hyperref}
\DeclareUnicodeCharacter{2212}{-} 

\usepackage{cancel}

\makeatletter
\def\@email#1#2{%
 \endgroup
 \patchcmd{\titleblock@produce}
  {\frontmatter@RRAPformat}
  {\frontmatter@RRAPformat{\produce@RRAP{*#1\href{mailto:#2}{#2}}}\frontmatter@RRAPformat}
  {}{}
}
\makeatother
\begin{document}

\preprint{AIP/123-QED}

\title[]{Multi-shot readout error benchmark of the nitrogen-vacancy center’s electronic qubit}

\author{Péter Boross}
    \affiliation{Qutility @ Faulhorn Labs, Budafoki út 91-93, H-1117 Budapest, Hungary}

\author{Domonkos Svastits}
    \affiliation{Qutility @ Faulhorn Labs, Budafoki út 91-93, H-1117 Budapest, Hungary}
    \affiliation{Department of Theoretical Physics, Institute of Physics, Budapest University of Technology and Economics, Műegyetem rkp. 3, H-1111 Budapest, Hungary}
    
\author{Győző Egri}
    \affiliation{Qutility @ Faulhorn Labs, Budafoki út 91-93, H-1117 Budapest, Hungary}

\author{András Pályi}
    \affiliation{Department of Theoretical Physics, Institute of Physics, Budapest University of Technology and Economics, Műegyetem rkp. 3, H-1111 Budapest, Hungary}
    \affiliation{HUN-REN-BME-BCE Quantum Technology Research Group, Műegyetem rkp. 3, H-1111 Budapest, Hungary}

\email{peter.boross@faulhornlabs.com, palyi.andras@ttk.bme.hu}

\date{\today}

\begin{abstract}
The ground-state electronic spin of a negatively charged nitrogen-vacancy center in diamond can be used for room-temperature experiments showing coherent qubit functionality.
At room temperature, photoluminescence-based qubit readout has a low single-shot fidelity; however, the populations of the qubit's two basis states can be inferred using multi-shot readout.
In this work, we calculate the dependence of the error of a multi-shot inference method on various parameters of the readout process. 
This multi-shot readout error scales as $\Delta/\sqrt{N}$, with $N$ being the number of shots, suggesting to use the coefficient $\Delta$ as a simple multi-shot readout error benchmark. 
Our calculation takes into account background photons, photon loss, and initialization error. Our model enables the identification of the readout error budget, i.e., the role various imperfections play in setting the readout error. Our results enable experimentalists and engineers to focus their efforts on those hardware improvements that yield the highest performance gain for multi-shot readout. 
\end{abstract}

\maketitle

\section{Introduction}

The ground-state electron configuration of a single, negatively charged nitrogen-vacancy center in diamond (NV)\cite{Gruber,DohertyReview} can be used as a quantum bit. 
This qubit, based on the electronic spin states of the NV, can be initialised, coherently controlled, and read out even at room temperature \cite{JelezkoPRL2004,HansonPRB2006}. 
Areas of application of such a qubit includes various subfields of quantum technology, e.g., quantum sensing \cite{TaylorNature2008,Balasubramanian,DoldeEfield,Hong}, quantum computing \cite{vanderSar,Taminiau,NeumannRegister}, as well as quantum technology education \cite{Sewani,YangYang}.

In quantum science and technology, single-shot qubit readout with low readout error is a valuable asset; in fact, for certain applications it is indispensable.
Unfortunately, at room temperature, photoluminescence-based readout of the NV's electronic spin qubit has a high single-shot error \cite{Hopper}, of the order of a few tens of percents.
Even with this limitation, it is possible to determine the populations of the qubit basis states with high precision, at the price of repeating the same experiment many times, and correspondingly, performing multi-shot readout.

This multi-shot approach to readout makes the hardware useful for a range of quantum information protocols, such as calibration, tomography, sensing, as well as certain quantum algorithms, where the inference of qubit-state populations or expectation values of qubit projections are sufficient. 
An example for such an algorithm is quantum simulation\cite{KimUtility}, where expectation values have to be inferred; a counterexample is Shor's algorithm, where specific bitstrings must be obtained at the end of the circuit in a single-shot manner\cite{Shor}.
Interestingly, multi-shot readout of the electron spin qubit can be exploited to perform high-fidelity single-shot readout of a nuclear spin in the vicinity of the NV\cite{NeumannNuclearSpinReadout, PhysRevLett.110.060502, PhysRevLett.118.150504, jaeger2024modelingquantumvolumeusing}.

The quality of single-shot readout of the NV's electronic qubit is often characterized \cite{Hopper} by the \emph{readout error} or \emph{readout infidelity} $\epsilon$.
One way to define it is $\epsilon = (\epsilon_{0}+\epsilon_1)/2$, where $\epsilon_0$ ($\epsilon_1$) is the probability of erroneously inferring a qubit value of 1 (0) when the qubit was in state 0 (1).
This single-shot readout error is one of the primary qubit-level benchmarks of quantum computers.

In multi-shot readout\cite{RyanTomography,DAnjou}, the goal is to infer the populations of the qubit basis states, instead of inferring a binary qubit value as in single-shot readout.
In other words, multi-shot readout is a parameter estimation problem for the continuous parameter $p \in [0,1]$ representing the population of qubit state $\ket{0}$.
Ideally, the error of the parameter estimation should decrease and converge to zero as the number of shots is increased to infinity. 

\begin{figure*}
\centering
\includegraphics[width=1\linewidth]{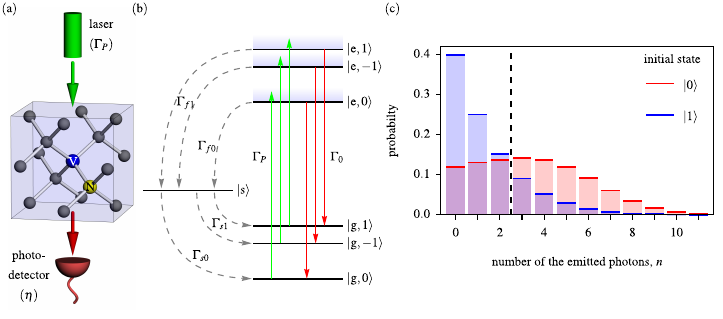}
\caption{\label{fig:setup}Optical readout of the electronic qubit of a negatively charged nitrogen-vacancy center (NV). 
(a) Qubit readout is based on counting the red photoluminescence photons emitted by the NV upon excitation with a green laser pulse, using a photodetector. 
Probability $\eta$ is the ratio of the detected photons and the emitted photons.
(b) Level diagram of a rate-equation model of the photoluminescence dynamics. 
The computational basis states of the qubit are the two lowermost levels, $\ket{0} = \ket{\text{g},0}$ and $\ket{1} = \ket{\text{g},-1}$.
(c) Red (blue): Probability mass function of the number of emitted photons upon readout, when the qubit state before the readout is $\ket{0}$ ($\ket{1}$).
Single-shot readout can be based on, e.g., the maximum likelihood inference rule, corresponding to the dashed black line serving as the separator between the inferred bit values.
The overlap of the two PMFs is strong, hence single-shot readout error probability is high, hence often multi-shot readout is used. The rates used to calculate the PMFs are 
    $\Gamma_P = 630$~MHz, 
    $\Gamma_0 = 63$~MHz, 
    $\Gamma_{f0} = 12$~MHz, 
    $\Gamma_{f1} = 80$~MHz, 
    $\Gamma_{s1}=2.4$~MHz, $\Gamma_{s0}=3.3$~MHz. \cite{Robledo_2011}
}
\end{figure*}

In this work, we propose a numerical readout error benchmark $\Delta$ for multi-shot readout, based on an adequately chosen estimator whose mean squared error decreases with the shot number $N$ as $\Delta^2/N$.
We show that the best multi-shot readout corresponds to $\Delta = \sqrt{2/3} \approx 0.816$; otherwise it holds that $\Delta > \sqrt{2/3}$.
We use a dynamical photoluminescence model for NVs to evaluate the dependence of this benchmark $\Delta$ on various parameters of an NV setup (see Fig.~\ref{fig:setup}(a)), including the dependence on the photon detection efficiency $\eta$, the background photon flux $\lambda$, and the initialization error probability $q$. 
We expect that the multi-shot readout error benchmark $\Delta$ could become a key indicator for readout quality for quantum computer prototypes relying on multi-shot readout, including NV-based devices. 

The rest of this paper is structured as follows. 
In Sec.~\ref{sec:multi-shot-readout-error}, we introduce multi-shot readout error in a general context, without direct reference to a specific qubit realization, readout apparatus, etc. 
In Sec.~\ref{sec:nv}, we specify a rate-equation model for the optical readout of the electronic qubit of an NV, compute photon-number distributions from this model, and use those to evaluate the parameter dependence of multi-shot readout error of NVs.
Here, we account for imperfect photon detection, a nonzero background photon flux, as well as imperfect qubit initialization. 
In Sec.~\ref{sec:discussion}, we provide a detailed discussion of our results, and we conclude in Sec.~\ref{sec:conclusions}.

\section{Multi-shot readout error}
\label{sec:multi-shot-readout-error}

Multi-shot qubit readout is an estimation task which can be formulated in quantum information theory language as follows.
Denote the computational basis states as $\ket{0}$ and $\ket{1}$.
We assume that we receive $N$ identical copies of a qubit state $\rho$ with ground-state population $\rho_{00} = p$ that is unknown to us.
For example, this qubit state can be a pure state, $\rho = \ket{\psi}\bra{\psi}$ with
$\ket{\psi} = \alpha \ket{0} + \beta \ket{1}$ and
$|\alpha|^2 + |\beta|^2 = 1$; in this case, 
we have $p = |\alpha|^2$.
We use the readout apparatus for all $N$ copies. 
Readout for a single copy (i.e., a \emph{single shot}) provides, in a probabilistic fashion, data that is determined by the qubit state as well as the details of the readout procedure.
We assume that for each shot, the data provided by the apparatus is a nonnegative integer $n$.
In the case of NVs, read out by photoluminescence, $n$ represents the number of detected photons by the photodetector in a single shot.
In general, $n$ is a random variable. 
We assume that its probability mass function (PMF) is the population-weighted sum $P(n) = p P_0(n) + (1-p) P_1(n)$
of two PMFs $P_0(n)$ and $P_1(n)$ that corresponds to the two computational basis states. 

The probability $p$ can be expressed using the expectation value 
of $\hat{\sigma}_z =\ket{0}\bra{0}- \ket{1}\bra{1}$, 
that is, $p = (1+z)/2$ with 
$z= \langle  \hat{\sigma}_z\rangle = \mathrm{Tr}(\hat{\rho} \hat{\sigma}_z)$. 
The task in multi-shot readout is to infer $z$ from the sequence of photon counts $\boldsymbol{n} = (n_1,n_2,\dots n_N)$ obtained in the $N$ shots.
Finding and characterizing solutions of such inference tasks is the subject of estimation theory; we will use that conceptual framework here, building our analysis on Chapter 9 of Ref.~\onlinecite{Kay}.

The task of inferring $z$, defined above, is carried out using an \emph{estimator}: a function $\hat{z}$ that maps the data $\boldsymbol{n}$ to a parameter value.
A natural estimator can be found via the following simple line of thought.
For a given parameter value $z$, and photon detection PMFs $P_0(n)$ and $P_1(n)$, the photon-number expectation value in a single shot reads
\begin{equation}
\label{eq:nbar}
    \mu_z = \frac{1+z}{2} \mu_0 + \frac{1-z}{2} \mu_1,
\end{equation}
where $\mu_0 = \sum_{n=0}^{\infty} n P_0(n)$ is the single-shot photon-number expectation value for qubit state $\ket{0}$, and $\mu_1$ is the corresponding single-shot photon-number expectation value for state $\ket{1}$.
Without the loss of generality, we will assume that $\mu_1 < \mu_0$.
We can express $z$ from Eq.~\eqref{eq:nbar} as follows:
\begin{equation}
\label{eq:pfromnbar}
z = \frac{\mu_z - (\mu_0 + \mu_1)/2}{(\mu_0-\mu_1)/2}.
\end{equation}

We obtain our estimator from Eq.~\eqref{eq:pfromnbar}, by replacing the single-shot photon-number expectation value $\mu_z$ by the empirical average single-shot photon count from $N$ shots, that is, 
\begin{equation}
\bar{n} = \frac{1}{N} \sum_{j=1}^N n_j.
\end{equation}
This yields
\begin{equation}
\label{eq:estimator}
\hat{z}(\boldsymbol{n}) := 
\frac{\bar{n} - (\mu_0 + \mu_1)/2}{(\mu_0 - \mu_1)/2}.
\end{equation}
Note that the value of the estimator $\hat{z}$ depends on the data $\boldsymbol{n}$ only via the empirical average single-shot photon count $\bar{n}$, hence we can use the notation $\hat{z}(\bar{n})$ in what follows. Alternatively, the same estimator can be derived from the general principle called `method of moments', see Chapter 9.3 of Ref.~\onlinecite{Kay}.
Note also that similar, so-called \emph{soft-average} estimators have been studied in the case of qubit readout via continuous measurement data\cite{RyanTomography,DAnjou}.

We characterize the accuracy of the estimator $\hat{z}$ defined in Eq.~\eqref{eq:estimator}  by its mean squared error:
\begin{equation}
\mathrm{MSE}(z,N) = \sum_{\bar{n} = 0}^\infty
P_\mathrm{multi}(\bar n;N,z) \left[\hat{z}(\bar n) - z \right]^2,
\label{eq:mse-def}
\end{equation}
where $P_\mathrm{multi}(\bar n;N,z)$ is the PMF of $\bar{n}$, assuming we take $N$ shots and that the parameter value is $z$. 
Note that the sum in Eq.~\eqref{eq:mse-def} runs for integer multiples of $1/N$.
In App.~\ref{app:mse-calculations}, we derive from Eq.~\eqref{eq:mse-def} that
\begin{subequations}
\label{eq:mseresult}
\begin{eqnarray}
    \mathrm{MSE}(z,N) &=& \frac{1}{N}\left( \frac{4(A + Bz+ Cz^2)}{(\mu_0-\mu_1)^2} \right),\\
    A &=& \left(\frac{\mu_0-\mu_1}{2}\right)^2 + \frac{\sigma_0^2 + \sigma_1^2}{2}, \\
    B &=& \frac{\sigma_0^2-\sigma_1^2}{2}, \\
    C &=& -\left(\frac{\mu_0-\mu_1}{2}\right)^2,
\end{eqnarray}
\end{subequations}
where $\sigma_0^2$ and $\sigma_1^2$ are the variances of the distributions $P_0(n)$ and $P_1(n)$. Eq.~\eqref{eq:mseresult} shows that the mean squared error of the estimator $\hat{z}$ is inversely proportional to the shot number $N$, and hence converges to zero in the limit $N\to \infty$.

We note that the estimator $\hat{z}$ defined in Eq.~\eqref{eq:estimator} is unbiased.
This follows from a straightforward calculation of its bias $\sum_{\bar{n}=0}^{\infty} P_\mathrm{multi}(\bar{n};N,z) \left[\hat{z}(\bar{n})-z \right]$.

To characterize multi-shot readout error with a single number, we average our result for the mean squared error Eq.~\eqref{eq:mseresult} over the values of the parameter $z$, assuming that $z$ is uniformly distributed in $[-1,1]$; that is, its prior is the constant function $P_\textrm{prior}(z) = 1/2$ on that interval.
Note that this averaging is equivalent to averaging for pure states according to the Haar measure.
The averaging yields:
\begin{align}
\label{eq:avgmseresult}
    \overline{\mathrm{MSE}}(N) = \frac{1}{2}\int_{-1}^1 dz \, \mathrm{MSE}(z,N) = \frac{\Delta^2}{N},
\end{align}
where 
\begin{align}
\label{eq:deltaplain}
    \Delta = \sqrt{\frac{2(\sigma_0^2+ \sigma_1^2)}{(\mu_0-\mu_1)^2} + \frac{2}{3}}.
\end{align}
We will refer to $\overline{\mathrm{MSE}}(N)$ as the \emph{multi-shot readout error} or \emph{$N$-shot readout error}, and $\Delta$ as the \emph{multi-shot readout error benchmark}.
According to Eq.~\eqref{eq:avgmseresult}, the multi-shot readout error error also shows $1/N$ dependence, inherited from Eq.~\eqref{eq:mseresult}.

The quantity $\Delta$ is a benchmark in the following sense. 
If qubit $A$ has a smaller $\Delta$ value than qubit $B$, that signals that multi-shot readout on qubit $A$ is more efficient than on qubit $B$: to achieve a certain small error $\overline{\mathrm{MSE}}(N) < \epsilon$, the user has to take less shots on qubit $A$ than on qubit $B$.

A natural question is: what is the smallest achievable value of the multi-shot readout error $\Delta$?
The first term under the square root in Eq.~\eqref{eq:deltaplain} is non-negative, which implies that
\begin{equation}
    \label{eq:fundamentallowerbound}
    \Delta \geq \sqrt{\frac{2}{3}} \approx 0.816,
\end{equation}
which is the fundamental lower bound on the multi-shot readout error benchmark $\Delta$.

\section{Multi-shot readout error model of the NV electron spin qubit at room temperature}
\label{sec:nv}

In this section, we evaluate the multi-shot readout error benchmark $\Delta$ of Eq.~\eqref{eq:avgmseresult} for NV-based spin qubits.
In particular, we focus on the optical readout of the electronic ground-state spin sublevels of the NV, and room-temperature operation where single-shot optical readout has low fidelity. 
We compute and illustrate how the multi-shot readout error is affected by imperfect photon detection, background photons, and imperfect qubit initialization. 

\subsection{Optical readout of an NV electronic qubit: setup and model}

In a standard experimental situation, a single NV is used, which is subject to a magnetic field that is aligned with the NV axis. 
The magnetic field energetically separates the three spin sublevels of the electronic ground state, such that a qubit is formed by two of these sublevels, say, $\ket{\mathrm{g},0} \equiv \ket{0}$ and $\ket{\mathrm{g},-1} \equiv \ket{1}$.
Here, integers in $\ket{\mathrm{g},0}$ and $\ket{\mathrm{g},-1}$ refer to the magnetic quantum number $m_S$ of the spin sublevels.

To read out this qubit optically, a green laser pulse is sent to the sample, red photons are emitted via a photoluminescence process, and those photons are detected.
The setup is illustrated in Fig.~\ref{fig:setup}(a).
The number of detected photons carries information about the qubit state.
The two PMFs, describing the number of detected photons for qubit state $\ket{0}$ and qubit state $\ket{1}$, are different (as illustrated in Fig.~\ref{fig:setup}(c)), but at room temperature they are not distinct enough to provide high-fidelity single-shot readout.
Multi-shot readout resolves that problem, at least for quantum information protocols where the inference of qubit-state populations or expectation values of qubit projections are sufficient.

Here, we implement a 7-level model\cite{Panadero}
to quantitatively describe the photolouminescence dynamics that occurs in the readout process.
The usefulness of this model has been evidenced by comparing it to experiments, see Figs.~4a, 7, and 8 in Ref.~\onlinecite{Panadero}.
The seven levels are shown in Fig.~\ref{fig:setup}(b), and labeled with 
$\sigma \in S \equiv \{\ket{\text{g},0},\ket{\text{g},-1},\ket{\text{g},1},\ket{\text{e},0},\ket{\text{e},-1},\ket{\text{e},1},\ket{\text{s}}\}$.
The first three states correspond to the ground-state triplet often referred to as $^3A_2$, the second three states correspond to the excited-state triplet $^3E$, and the seventh state is an intermediate singlet state that enables non-radiative decay from the excited triplet to the ground-state triplet. We note that the effect of the misaligned magnetic field relative to the NV axis on the transition rates can be described as formulated in Ref. \onlinecite{Tetienne2012}.

This model enables the computation of the photon number PMFs $P_0(n)$ and $P_1(n)$ introduced and used in the previous section.
As we have shown there, from these PMFs we can evaluate the multi-shot readout error benchmark $\Delta$.
The model's input parameters are the laser pulse duration, the rates characterizing the various radiative and non-radiative transitions among the 7 energy levels, and the laser pulse intensity, represented by the rate $\Gamma_P$ in Fig.~\ref{fig:setup}(b).

The level diagram, the various transitions, and the corresponding rates, are shown in Fig.~\ref{fig:setup}(b).
A single photon is emitted in each of the three transitions denoted by the three red downward arrows, labeled by $\Gamma_0$.
To compute the photon number PMFs $P_0(n)$ and $P_1(n)$, we solve the photon-number-resolved rate equation that governs the dynamics, detailed in App.~\ref{app:rateequation}.
In this rate equation, the unknown is the infinite vector $\boldsymbol{p}(t)$, whose entry $p_{\sigma,n}(t)$ is the probability of occupying level $\sigma$ at time $t$ after the switch-on of the laser pulse, such that $n$ photoluminescence photons have been emitted since the switch-on of the laser.
That is, the normalization condition reads
$\sum_{\sigma \in S} \sum_{n=0}^\infty p_{\sigma,n}(t)= 1$ for all times $t$.

After solving the rate equation for initial probability vector $p_{\sigma,n}(t=0) = \delta_{\sigma,\ket{\mathrm{g},0}} \delta_{n,0}$, we compute the photon number PMF $P_0(n)$ as follows.
For simplicity, we assume that the laser pulse and the detection of photoluminescence photons occurs in the same time window from $t=0$ till $t=t_\mathrm{m}$.
Then, the photon number PMF $P_0(n)$ is expressed as
\begin{equation}
\label{eq:simulation2PMF}
P_0(n) = \sum_{\sigma \in S} p_{\sigma,n}(t_\mathrm{m}),
\end{equation}
and we obtain $P_1(n)$ from the same formula, using the solution evolving from the initial probability vector $p_{\sigma,n}(t=0) = \delta_{\sigma,\ket{\mathrm{g},-1}} \delta_{n,0}$.

We solve the infinite set of rate equations for a laser pulse of finite duration $t_\mathrm{m}$ by truncating the equation in photon number, keeping the entries of $\boldsymbol{p}$ for $n\in \{0,1,2,\dots, n_\mathrm{max}\}$ only, and checking for the convergence of the results as $n_\mathrm{max}$ is increased.

\subsection{Perfect photon detection}

The photon number PMFs are exemplified in Fig.~\ref{fig:setup}(c), where the red (blue) PMF corresponds to the qubit state $\ket{0}$ ($\ket{1}$).
These histograms are obtained using the rate values shown in the caption of Fig.~\ref{fig:MSE}, laser pulse duration $t_\mathrm{m} = 92$ ns, and assuming perfect photon detection, that is, $\eta = 1$. 
Photon detection efficiency is characterized by the parameter $0 \leq \eta \leq 1$, which is defined as the fraction of those photoluminescence photons that are detected by the detector. 
Knowing these histograms, one can do single-shot readout with binary inference based on the maximum likelihood principle: we infer $\ket{0}$ if $n>2$ and $\ket{1}$ otherwise. 
For the histograms shown in Fig.~\ref{fig:setup}(c), the errors of the corresponding single-shot readout are $\epsilon_0 = 0.39$, $\epsilon_1 =0.20 $, $\epsilon = 0.29 $.

Here, we focus on applications where the task is to infer $z$, i.e., the expectation value of $\hat{\sigma}_z$, and consider the multi-shot inference method described in Sec.~\ref{sec:multi-shot-readout-error}.
By computing the photon counting PMFs $P_0(n)$ and $P_1(n)$ from our simulations via Eq.~\eqref{eq:simulation2PMF}, and inserting their means and variances into Eq.~\eqref{eq:deltaplain}, we obtain the multi-shot readout error benchmark $\Delta$,
corresponding to the PMFs in Fig.~\ref{fig:setup}(c), as $\Delta \approx 2.1$.

\subsection{Imperfect photon detection}
\label{sec:imperfect}

So far, we have assumed perfect photon detection, $\eta =1$. In state-of-the-art experiments, a photon detection efficiency approaching unity has been achieved \cite{lubotzky2025}. However, this requires advanced techniques, therefore, we consider the case of imperfect photon collection, that is, $\eta < 1$.

So far, $P_0(n)$ denoted simultaneously the PMF of the number of emitted photons and the PMF of the number of detected photons. 
As we take into account $\eta < 1$ from now on, we need to distinguish these two PMFs; in what follows, we will use $P_0^\mathrm{e}(n)$ for the number of emitted photons, and keep $P_0(n)$ for the number of detected photons.
Similarly, $\mu_0^\mathrm{e}$, $\mu_1^\mathrm{e}$, 
$\sigma_0^\mathrm{e}$, 
$\sigma_1^\mathrm{e}$ refer to 
the properties of the PMFs of the number of emitted photons. 

The relation between the PMFs describing the emitted and detected photons is expressed as:
\begin{equation}
     P_{0/1}(n) = \sum_{i = n}^\infty P_{0/1}^\mathrm{e}(i)\cdot{i\choose n} \eta^n (1-\eta)^{(i-n)}.
     \label{eq:distr-photon-count-eff}
\end{equation}
This formula derives from the following simple consideration: we detect $n$ photons whenever the number of emitted photons $i$ is greater or equal than $n$, and $n$ of the $i$ emitted photons are detected.
Furthermore, we show in App.~\ref{app:mse-calculations} that the means and variances of the above PMFs are related as follows:
\begin{eqnarray}
    \mu_{0/1} &=& \eta \mu_{0/1}^\mathrm{e}, 
    \label{eq:mean-efficiency}\\
    \sigma_{0/1}^2 &=& (1-\eta)\eta \mu_{0/1}^\mathrm{e} + \eta^2 (\sigma_{0/1}^\mathrm{e})^2.
    \label{eq:variance-efficiency}
\end{eqnarray}

Substituting Eqs.~\eqref{eq:mean-efficiency} and \eqref{eq:variance-efficiency} into Eq.~\eqref{eq:deltaplain}, we obtain the multi-shot readout error benchmark as
\begin{equation}
    \Delta =\left(\frac{2(1-\eta) (\mu_0^\mathrm{e}+\mu_1^\mathrm{e})}{\eta(\mu_0^\mathrm{e}-\mu_1^\mathrm{e})^2}+\frac{2\left[(\sigma_0^\mathrm{e})^2 + (\sigma_1^\mathrm{e})^2\right]}{(\mu_0^\mathrm{e}-\mu_1^\mathrm{e})^2} + \frac{2}{3}\right)^{1/2}.
    \label{eq:avgMSE-efficiency}
\end{equation}
Note that the dependence of the multi-shot readout error $\Delta$ on the photon detection efficiency $\eta$ is through the first term on the right hand side of Eq.~\eqref{eq:avgMSE-efficiency}, whereas the means and variances on the right hand side of Eq.~\eqref{eq:avgMSE-efficiency} depend on the transition rates and the measurement duration $t_\mathrm{m}$.

\begin{figure*}
    \centering
    \includegraphics[width=\linewidth]{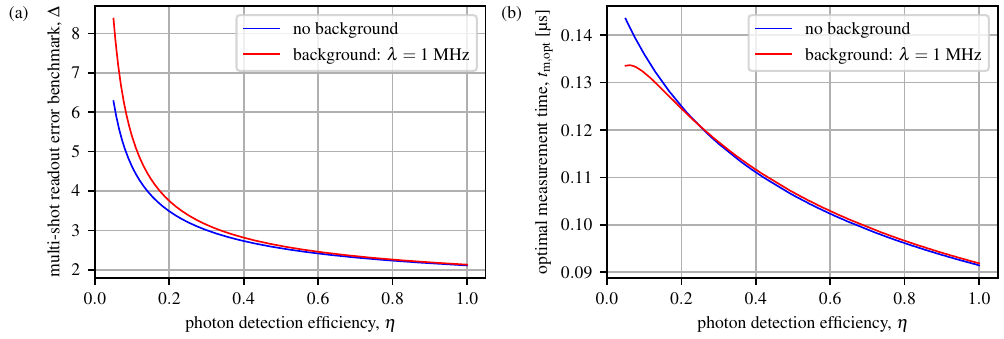}
    \caption{
    \label{fig:MSE}Multi-shot readout error $\Delta$ for an NV electronic qubit, affected by the photon detection efficiency and background photons. 
    Uncertainty of the multi-shot estimate of the expectation value of the qubit's $\hat{\sigma}_z$ varies with the number of shots $N$ as $\Delta/\sqrt{N}$.
    (a) Multi-shot readout error $\Delta$ as function of photon detection efficiency $\eta$, optimized for the readout laser pulse duration. 
    Blue: without background photons.
    Red: with background photon flux $\lambda = 1$ MHz.
    (b) Optimal values of the readout laser pulse duration. For the values of the rates, see caption of Fig.~\ref{fig:setup}.
    }
\end{figure*}

This dependence $\Delta (\eta)$ is illustrated in Fig.~\ref{fig:MSE}.
We set the rate parameters of our simulation to the values in the caption of Fig.~\ref{fig:MSE}, and for each value of $\eta$, we optimize (that is, minimize) the multi-shot readout error with respect to the measurement duration $t_\textrm{m}$.
The optimization was carried out numerically, by evaluating $\Delta$ at 5000 equidistant values of $t_\mathrm{m}$ between between $0$ and $200 \, \mathrm{ns}$. The dependence of $\Delta$ on $t_\mathrm{m}$ is discussed in App.~\ref{app:meas-time-dependence}.
The optimal $\Delta$ values are shown as the function of the photon detection efficiency $\eta$ in Fig.~\ref{fig:MSE}(a) as the blue line, and the optimized measurement durations $t_\mathrm{m,opt}$ are  shown in Fig.~\ref{fig:MSE}(b) as the blue line.

Most importantly, Fig.~\ref{fig:MSE}(a)  conveys and quantifies the message that increasing the photon detection efficiency suppresses the multi-shot readout error.
According to our approach, the multi-shot readout error for  photon detection efficiency $\eta = 0.1$ (see, e.g., Ref.~\onlinecite{WanParabolicReflector}) is $\Delta \approx 4.6$.
This number can be used to compute the number of shots required to obtain an estimation of the $\hat{\sigma}_z$ expectation value with a certain error target: e.g., to reach an error of $1\%$, one needs to run $N = \left(\Delta/0.01\right)^2 \approx 2 \times 10^5$ shots. 
Another important message of Fig.~\ref{fig:MSE} is that the optimal value of $\Delta$ for perfect photon detection $\eta = 1$ is $\Delta \approx 2.1$, which is still significantly greater than the fundamental lower bound $\sqrt{2/3} \approx 0.816$ (see Eq.~\eqref{eq:fundamentallowerbound}).
This result also conveys the message that in order to approach the fundamental lower bound, other measures have to be taken besides increasing $\eta$ (see Sec.~\ref{sec:discussion}).

\subsection{Background photons}

So far, we assumed that all photons detected by the detector arrive from the qubit. 
However, background radiation can also contribute to the detector's photon count, and hence modify the PMFs $P_0(n)$ and $P_1(n)$ and, in turn, the multi-shot readout error. 
We assume that the number of background photons $n_b$ detected during the measurement time $t_\mathrm{m}$ are described by a Poisson process:
\begin{equation}
    n_b \sim \mathrm{Poi}(\lambda t_\mathrm{m}),
\end{equation}
where the \emph{background photon flux} $\lambda$ is the parameter of the Poisson process. 
The unit of the background photon flux $\lambda$ is Hz or s$^{-1}$.

The number of detected photons $n_{0/1}$ is the sum of the number of photons that were detected \emph{and} arrived from the NV-center ($n_{0/1}^\mathrm{NV}$) and the number of detected background photons ($n_b$):
\begin{equation}
    n_{0/1} = n_b + n_{0/1}^\mathrm{NV}.
    \label{eq:photon-count-background}
\end{equation}
Substitution of the variances and expectation values of these random variables into Eq.~\eqref{eq:deltaplain} yields
\begin{equation}
    \Delta = \left(\frac{2\left[\left(\sigma_0^\mathrm{NV}\right)^2 + \left(\sigma_1^\mathrm{NV}\right)^2 + 2\lambda t_m\right]}{\left({\mu}_0^\mathrm{NV}-{\mu}_1^\mathrm{NV}\right)^2} + \frac{2}{3}\right)^{1/2}.
    \label{eq:delta-background}
\end{equation}
Note that in the previous section, background photons were not considered, so all detected photons originated from the NV center. Consequently, $\mu_{0/1}^\mathrm{NV}$ and $\left(\sigma_{0/1}^\mathrm{NV}\right)^2$ are determined from the emitted photon statistics in the same way as $\mu_{0/1}$ and $\sigma_{0/1}^2$ in Eqs.~\eqref{eq:mean-efficiency} and \eqref{eq:variance-efficiency}, respectively. 

The effect of the background photons on the multi-shot readout error $\Delta$ is shown with the red line in Fig.~\ref{fig:MSE}, with a background photon flux $\lambda = 1$ MHz, which is an order of magnitude larger than in state-of-the-art experiments \cite{Panadero}. We used this larger value to better illustrate the effect of background photons in our plots. Fig.~\ref{fig:MSE}(a) shows that background photons increase the multi-shot readout error, and that this background-induced error component is suppressed for increasing photon detection efficiency.
Note that the time dependence of the extra term $2\lambda t_\mathrm{m}/({\mu}_0^\mathrm{NV}-{\mu}_1^\mathrm{NV})^2$ in Eq.~\eqref{eq:delta-background} is different from the time dependence of the other terms, therefore, the optimal measurement duration is changed when background photons are taken into account.
This is seen as the difference between the blue and red lines in Fig.~\ref{fig:MSE}(b). 

Note that in the above result, we have assumed that the background photon flux is independent of the photon detection efficiency $\eta$.
This is in fact a realistic scenario whenever the detector itself is close to perfect, and hence, $\eta$ represents the efficiency of directing the photoluminescence photons onto the detector.

\subsection{Imperfect initialization}
\label{app:imperfectinit}

In this section, we describe how imperfect initialization increases the multi-shot readout error. 
Imperfect initialization is a standard scenario for a room-temperature electronic NV qubit\cite{JelezkoPRL2004,Hopper}. 
In our approach, initialization error is characterized by the error probability $0 \leq q < 1/2$.
We assume that instead of the computational basis state $\ket{0}$, a mixture $\hat{\rho}_0 = (1-q)\ket{0}\bra{0}+ q\ket{1}\bra{1}$ of the two computational basis states is initialized.
Other states are accessed via unitary rotations of $\hat{\rho}_0$; for example, instead of the state $\ket{1}$, the state $\hat{\rho}_1=q\ket{0}\bra{0}+ (1-q)\ket{1}\bra{1}$ can be created from $\hat{\rho}_0$ by an X gate.

In this section, we regard the device as a single-qubit quantum computer.
Then, its functionality is not corrupted by initialization error, in the following sense.
The attainable states are not the pure states on the qubit Bloch sphere, but mixed states on a sphere of radius $1-2q$ inside the Bloch sphere.
However, one can reinterpret this inner sphere as the Bloch sphere, and instead of the true $\hat{\sigma}_z$ expectation value for a given qubit state $\hat{\rho}$, one can aim to estimate $z = \mathrm{Tr}(\hat{\rho} \hat{\sigma}_z)/(1-2q)$ from a multi-shot experiment;
the quantity $z$ would be the $\hat{\sigma}_z$ expectation value if the initalization was perfect.
In this context, the error of the estimate of $z$ increases with increasing initalization error $q$, as we show below.

The number of collected photons $n_0$ and $n_1$ corresponding to states $\hat{\rho}_0$ and $\hat{\rho}_1$ are mixtures of the random variables $n_{\ket{0}}$ and $n_{\ket{1}}$ corresponding to the actual computational basis states $\ket{0}$ and $\ket{1}$, that is their PMFs are
\begin{equation}
    P_{0/1}(i) \equiv P(n_{0/1}=i) = (1-q)P\left(n_{\ket{0/1}}=i\right) + q P\left(n_{\ket{1/0}}=i\right).
\end{equation}
The expectation values and variances of these random variables are
\begin{eqnarray}
    \label{eq:muinit}
    \mu_{0/1} &=& (1-q)\mu_{\ket{0/1}} + q \mu_{\ket{1/0}},\\
    \label{eq:sigmainit}
    \sigma_{0/1}^2 &=& \sigma_{\ket{0/1}}^2 + q(1-q) \left( \mu_{\ket{0/1}}-\mu_{\ket{1/0}} \right)^2 \nonumber \\
    &+& q \left( \sigma_{\ket{1/0}}^2 - \sigma_{\ket{0/1}}^2\right).
\end{eqnarray}
We derive the relations in Eqs.~\eqref{eq:muinit} and \eqref{eq:sigmainit} in App.~\ref{app:mse-calculations}.

Substituting Eqs.~\eqref{eq:muinit} and \eqref{eq:sigmainit} into Eq.~\eqref{eq:deltaplain}, we obtain: 
\begin{equation}
    \Delta = \left(\frac{2\left(\sigma_{\ket{0}}^2 + \sigma_{\ket{1}}^2\right)}{\left(\mu_{\ket{0}}-\mu_{\ket{1}}\right)^2(1-2q)^2} + \frac{4q(1-q)}{(1-2q)^2} + \frac{2}{3}\right)^{1/2}.
\end{equation}
The variances $\sigma_{\ket{0}}^2$ and $\sigma_{\ket{1}}^2$ and expectation values $\mu_{\ket{0}}$ and $\mu_{\ket{1}}$ can be further expressed by the properties of the emitted photon distributions, the photon detection efficiency $\eta$, and the background photon flux $\lambda$, using Eqs.~\eqref{eq:mean-efficiency}, \eqref{eq:variance-efficiency} and \eqref{eq:photon-count-background}, yielding the following general formula for the multi-shot readout error benchmark:
\begin{widetext}
\begin{equation}
    \label{eq:deltageneral}
    \Delta = \left(\frac{2\left[(1-\eta)\eta\left({\mu}_{\ket{0}}^\mathrm{e}+{\mu}_{\ket{1}}^\mathrm{e}\right)+\eta^2\left(\left(\sigma_{\ket{0}}^\mathrm{e}\right)^2 + \left(\sigma_{\ket{1}}^\mathrm{e}\right)^2\right) + 2\lambda t_\mathrm{m}\right]}{\eta^2({\mu}_{\ket{0}}^\mathrm{e}-{\mu}_{\ket{1}}^\mathrm{e})^2(1-2q)^2} + \frac{4q(1-q)}{(1-2q)^2} + \frac{2}{3}\right)^{1/2}.
\end{equation}
\end{widetext}
Note that the means and variances on the right hand side do depend on the laser pulse duration $t_\mathrm{m}$.

\begin{figure}
    \centering
    \includegraphics[width=\linewidth]{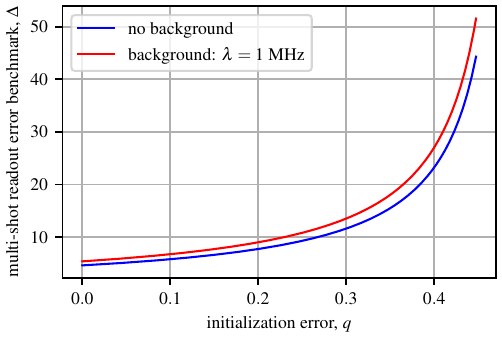}
    \caption{Multi-shot readout error $\Delta$ as a function of initialization error. 
    Parameters: 
    $\eta = 0.1$; see caption of Fig.~\ref{fig:setup} for rate values. 
    Blue: no background photons, $\lambda = 0$. 
    Red: background photon flux $\lambda=1$~MHz.
    Readout pulse durations are the optimal values indicated in Fig.~\ref{fig:MSE}(b) for $\eta = 0.1$; that is, 
    $t_\mathrm{m} = 136$ ns ($t_\mathrm{m} = 132$ ns) for the blue (red) data set.}
    \label{fig:delta-q}
\end{figure}

The dependence of the multi-shot readout error benchmark on the initialization error is shown Fig.~\ref{fig:delta-q}.
The readout error is optimized with respect to the laser pulse duration (see caption for duration values).
Note that the optimal laser pulse duration is independent of the initialization error $q$.
This follows from Eq.~\eqref{eq:deltageneral}, where only the first fraction depends on the laser pulse duration via the means, variances, and the $2\lambda t_\mathrm{m}$ term.
Figure \ref{fig:delta-q} shows that both without (blue) and with (red) background photons, an increasing initialization error increases the multi-shot readout error.

To conclude, in this section we have computed the multi-shot readout error benchmark $\Delta$ from a rate-equation model of the NV readout photocycle, taking into account imperfect photon detection, background photons, and initialization error.

\section{Discussion}
\label{sec:discussion}

\subsection{Experimental evaluation of the multi-shot readout error benchmark}

In Sec.~\ref{sec:nv}, we evaluated the multi-shot readout error benchmark $\Delta$ from simulations. Here, we outline how to evaluate it from experimental data.
The first step is to calibrate initialization, readout, as well as an $X$ gate, e.g., based on electronic Rabi oscillations induced by ac magnetic field pulses. 
The second step is to infer the single-shot photon-counting PMFs $P_0(n)$ ($P_1(n)$), by doing many shots, each shot consisting of initialization and readout (initalization, X gate, and readout).
The third, final step is to compute the means and variances of these PMFs, and inserting those into Eq.~\eqref{eq:deltaplain} to obtain the multi-shot readout error benchmark.

\subsection{Applications of the multi-shot readout error benchmark}

NVs are used broadly in quantum education, science and technology, as a popular and relatively simple experimental tool, and  also as a platform for developing commercial hardware.
Therefore, benchmarks characterizing the components of NV-based technology, such as the multi-shot readout error benchmark $\Delta$ we propose, can play central roles in research, development, and commercialization; we list a few examples in what follows.

\emph{Evaluation of either the N-shot readout error or the required number of shots.}
The most natural use case of the multi-shot readout error benchmark $\Delta$ is to evaluate the $N$-shot readout error
\begin{equation}
\label{eq:delta-vs-N}
    \delta = \Delta/\sqrt{N},
\end{equation}
given the number of shots $N$.
Take, for example, $\Delta \approx 4.6$, which is the case for the blue line at $\eta = 0.1$ in Fig.~\ref{fig:MSE}(a) (no background photons, perfect initialization). 
In that case, in an experiment of $N=10^5$ shots, the $N$-shot error of estimating $z$ is $\delta \approx 0.0145$.

Alternatively, the user can set a target error, e.g., $\delta=0.01$, and ask how many shots do we need to reach that error?
Inverting the relation in Eq.~\eqref{eq:delta-vs-N}, the answer is expressed as: 
\begin{equation}
    N = (\Delta / \delta)^2,
\end{equation}
which yields $N\approx 2 \times 10^5$ for $\Delta = 4.6$.

This description of the multi-shot readout error is of key importance for quantum algorithms aiming to determine expectation values of operators.
Here, we exemplify and illustrate such an application through the quantum simulation experiment of Ref.~\onlinecite{KimUtility}.
This experiment used a quantum computer based on superconducting qubits and high-fidelity single-shot readout. The goal of the experiment was to quantum-simulate the time evolution of Pauli expectation values in a Floquet-Ising model. A natural question is how to adapt that experiment to a room-temperature NV-based setup, where the qubit register has a single electronic qubit with low single-shot readout fidelity, and the rest of the register is formed by nuclear spins that cannot be read out directly?

A potential adapatation consists of the following steps: (1) Map the spins of the Floquet-Ising model onto the nuclear spins surrounding the NV. (2) Decide which Pauli operator’s expectation value we want to simulate. This can be a Pauli operator with weight 1, or with weight greater than 1. (3) Determine the Clifford unitary $U_\mathrm{C}$ that maps the chosen Pauli to the $\hat{\sigma}_z$ of the NV electronic spin, and compile it to native gates. (4) Run $N$ shots of the quantum simulation, i.e., of the following sequence: (i) initialisation, (ii) quantum simulation of the Floquet-Ising dynamics, (iii) execution of $U_\mathrm{C}$, (iv) readout of the NV electronic spin. From the $N$ shots, the expectation value of the chosen Pauli operator is inferred using Eq. (4) of our manuscript. 

For this experiment, it is critical to understand control the error of the inference of the Pauli expectation value. An important component of that is the multi-shot readout error we describe in our work. Even if steps (i), (ii), (iii) are perfect, step (iv) will imply a nonzero error for the inferred expectation value. This error component can be mitigated by an increased number of shots, and the efficiency of this mitigation is quantified by the multi-shot readout error benchmark $\Delta$.

\emph{Evaluation and optimization of the hardware.}
Experimentalists can use the benchmark to optimize experimental parameters to maximize multi-shot readout performance, that is, minimize $\Delta$.
This we exemplified by the results shown in Figs.~\ref{fig:MSE} and \ref{fig:delta-q}, where $\Delta$ was minimized over the laser pulse duration parameter. 
Further parameters that could be tuned and hence used for minimizing $\Delta$ are the laser intensity, magnetic field, electric field, mechanical strain, etc, which all have an effect of the transition rates governing the photocycle dynamics.
Besides the \emph{in situ} tuning of certain parameters, hardware optimization often entails replacing building blocks (e.g., lasers, passive optical elements, detectors, etc) of the setup; evaluating the benchmark quantifies the resulting improvement in terms of multi-shot readout, facilitating the optimization process. 

\emph{Cross-platform comparison.}
In this work, we express the multi-shot readout error benchmark $\Delta$ from the photon number PMFs. 
This is readily generalizable to other qubits which are read out by photon counting. 
Further, this concept is generalizable to cases where single-shot data comes in the form of real numbers, or vectors of real numbers. 
Hence, $\Delta$ (and its generalizations) can be considered as a cross-platform benchmark.

\emph{Use in commercialization and marketing.}
In commercial applications of NV centers, including quantum computer prototypes, the multi-shot readout error benchmark provides a metric to evaluate, compare, and market product performance.

\emph{Guiding hardware optimization and development.}
Our result in Eq.~\eqref{eq:deltageneral} expresses multi-shot readout error in terms of parameters that characterize certain `imperfections' of the setup: imperfect photon detection, background photons, and initialization error. 
This result is especially helpful for \emph{prioritizing} between different types of hardware improvements. 
In hardware development, one can often estimate the costs of improving certain components of the hardware with a certain amount.
For example, assume that improving the photon detection efficiency from a smaller value $\eta_1$ to a higher value $\eta_2$ has the same cost as improving initialization by decreasing the initialization error of a higher value $q_1$ to a lower value $q_2$.
In this case, Eq.~\eqref{eq:deltageneral} can be used to compute the corresponding improvement in terms of $\Delta$, and the change implying the greater improvement can be prioritized.

\subsection{Relation of the multi-shot readout error to the single-shot signal-to-noise ratio}

Ref.~\onlinecite{Hopper} collects various figures of merit quantifying the performance of optical readout of NVs.
A central figure of merit is the single-shot signal-to-noise ratio (SNR).
The relations of the SNR to further readout metrics are listed in Table 1 of Ref.~\onlinecite{Hopper}.

Here, we argue that the multi-shot readout error $\Delta$, the readout benchmark we have defined in this work, has a simple, monotonous relation with the SNR.
In our case, where the measured signal is the number of detected photons, SNR is defined as
\begin{equation}
    \mathrm{SNR} = \frac{\mu_0- \mu_1}{\sqrt{\sigma_0^2 + \sigma_1^2}}.
    \label{eq:SNR}
\end{equation}
Comparison of Eqs.~\eqref{eq:SNR} and \eqref{eq:deltaplain} yields that the multi-shot readout error benchmark $\Delta$ is related to the SNR as 
\begin{equation}
    \Delta = \sqrt{\frac{2}{\mathrm{SNR}^2}+ \frac{2}{3}}.
    \label{eq:delta-snr}
\end{equation}
This results highlights that $\Delta$ is a monotonic function of SNR; this implies that optimizing (maximizing) SNR by fine-tuning the control parameters of the readout simultaneously optimizes (minimizes) the multi-shot readout error benchmark $\Delta$ as well.

The SNR is commonly measured experimentally, therefore, using Eq.~\eqref{eq:delta-snr}, we can calculate typical values of the multi-shot readout error benchmark $\Delta$. We do this for two examples. (1) In Ref.~\onlinecite{Neumann2012Dissertation}, $\mathrm{SNR} \approx 0.1$ is reported, which translates to $\Delta \approx 14$. (2) From the photon count histograms shown in Fig. 3A of Ref.~\onlinecite{NeumannNuclearSpinReadout}, we estimate the single-shot SNR to be $\mathrm{SNR}\approx 0.06$, which translates to $\Delta\approx 23$.

\subsection{Further remarks}

\emph{Improvements in room-temperature readout of the NV's electronic qubit.}
A natural way to improve NV readout is to enhance the photon-collection efficiency, e.g., by embedding the NV in engineered photonic nanostructures\cite{Hopper,WanParabolicReflector,lubotzky2025,LouzhouLiBullseye2015}.
In our study, we assumed that the readout laser pulse is rectangular; however, the pulse shape could be optimized to lower the readout error, similar to how pulse-shape optimization has been exploited to improve initialization of the NV qubit\cite{Song}.
Further strategies\cite{Hopper} for improving readout includes radiative lifetime engineering, spin-to-charge conversion, nuclear-spin-assisted readout and the utilization of the time-of-arrival information of the photoluminescence photons\cite{Nakamura,Qian,Gupta}.

\emph{Role of ionization.}
The NV can switch between its negative and neutral charge state. These dynamics can be optically induced, for example by the laser pulse that is used for initialization or readout of the NV qubit.
In our context, this switching dynamics is an undesired event that complicates the processing of the data and degrades the quality of the readout.
Our rate-equation model can be extended to account for this dynamics and the corresponding changes in the quality of the readout, e.g., by using the model of Ref.~\onlinecite{Wirtitsch}. 
We also note that ionization can be also exploited for better readout as the mechanism enabling spin-to-charge conversion \cite{PhysRevLett.114.136402}.

\emph{Readout errors from gate miscalibration.} In NV-based quantum computer prototypes, the qubit register is usually formed by the electronic spin qubit and a few nearby nuclear spin qubits\cite{Robledo,Taminiau,Bradley}.
Optical readout cannot access the nuclear spins directly; nevertheless, it is possible to read out out a nuclear qubit indirectly, e.g., by performing a SWAP gate between that qubit and the electron spin, and then doing optical readout of the electronic qubit.
In such a scheme, the imperfection of the SWAP gate can introduce a readout error that is not mitigated by an increased shot number; in particular, this is the case when the SWAP is  poorly calibrated, and hence causes a systematic (as opposed to stochastic) error.

\section{Conclusions}
\label{sec:conclusions}

In conclusion, we have introduced a quantitative benchmark that characterizes multi-shot readout error, and evaluated how it depends on various system parameters when applied to the room-temperature readout of an electronic spin qubit of a negatively charged NV center.
We have quantified how the multi-shot readout error is impacted by imperfect photon detection, background photons, as well as imperfect qubit initialization.
We have also identified a simple relation between the multi-shot readout error and the signal-to-noise ratio. 
Room-temperature quantum computer prototypes based on point defects in semiconductors are already being commercialized, and we expect that our results will facilitate their characterization and further development.

\begin{acknowledgments}
We thank V.~Ivády, D.~Pataki, and our partners in the AI4QT project for useful and inspiring discussions.
This research was supported by the Hungarian Ministry of Culture and Innovation via the National Research, Development and Innovation Fund, through the AI4QT project Nr.~2020-1.2.3-EUREKA-2022-00029, 
via the EKÖP\_KDP-24-1-BME-2 funding scheme through Project Nr.~2024-2.1.2-EKÖP-KDP-2024-00005,
and by HUN-REN 3410107 (HUN-REN-BME-BCE Quantum Technology Research Group).
\end{acknowledgments}

\vspace{0.7cm}

\section*{Data Availability Statement}

The data that support the findings of this study are available from the corresponding author upon reasonable request.

\appendix

\section{Details of derivations for Section \ref{sec:multi-shot-readout-error}}
\label{app:mse-calculations}
Here, we detail the steps of three calculations: (1) deriving our result for the MSE in Eq.~\eqref{eq:mseresult} from its definition; (2) calculating the means and variances of the detected photon numbers from the emitted photon statistics for imperfect photon detection; and (3) determining how the means and variances of the detected photon numbers are affected by initialization error. The simplest way to do these calculations is using the law of total expectation and law of total variance, which state
\begin{align}
    \mathrm{E}(X) &= \sum_y P(y)\mathrm{E}(X|Y=y), 
    \label{eq:law-of-tot-exp} \\
    \mathrm{Var}(X) &= \sum_y P(y)\mathrm{Var}(X|Y=y) \nonumber \\
    &+ \sum_y P(y) \left(\mathrm{E}(X|Y=y) - \sum_z \mathrm{E}(X|Y=z)P(z) \right)^2,
    \label{eq:law-of-tot-var}
\end{align}
where $X$ and $Y$ are arbitrary random variables and the summations go over all possible values of $Y$.

\subsection{Calculating the MSE from the detected photon statistics}
Here, we show how our result for the MSE in Eq.~\eqref{eq:mseresult} is obtained. By substituting the estimator in Eq.~\eqref{eq:estimator} into the definition of the MSE in Eq.~\eqref{eq:mse-def}, we obtain
\begin{equation}
\begin{split}
    \mathrm{MSE}(z, N) &= \sum_{n=0}^\infty P_{\mathrm{multi}}(\bar{n}; N, z)\left(\frac{\bar{n}-\mu_z}{(\mu_0-\mu_1)/2}\right)^2=\\
    &=\frac{4}{(\mu_0-\mu_1)^2}\mathrm{Var}(\bar{n})=\frac{4}{(\mu_0-\mu_1)^2}\frac{\sigma_z^2}{N}.
    \label{eq:mse-var}
\end{split}
\end{equation}
Here, we used the notation $\sigma_z^2$ for the variance of the photon count for a single measurement of the state parameterized by $z$. At the last equation mark, we used that $\bar{n}$ is the arithmetic mean of the $N$ single-shot photon counts, which are independent identically distributed random variables with variance $\sigma_z^2$. 

Now, we express the variance $\sigma_z^2$ with the parameters $\sigma_0^2$, $\sigma_1^2$, $\mu_0$ and $\mu_1$.  
Let us denote the state being occupied by $s$, which is a random variable. It takes the values $s = 0$ and $s=1$ with probabilities $(1+z)/2$ and $(1-z)/2$ respectively. We can use the law of total variance, with $n$ and $s$ taking the roles of $X$ and $Y$ in Eq.~\eqref{eq:law-of-tot-var} respectively, which leads to 
\begin{equation}
\begin{split}
    \sigma_z^2 &= \frac{1+z}{2}\sigma_0^2 + \frac{1-z}{2}\sigma_1^2\\
    &+\frac{1+z}{2}(\mu_0-\mu_z)^2 + \frac{1-z}{2}(\mu_1-\mu_z)^2.
    \label{eq:sigma_z}
\end{split}
\end{equation}
Substituting Eqs.~\eqref{eq:nbar} and \eqref{eq:sigma_z} into Eq.~\eqref{eq:mse-var} and collecting the coefficients of different powers of $z$ leads to Eq.~\eqref{eq:mseresult}.

\subsection{Calculating properties of the detected photon statistics from the emitted photon statistics}
Here, we derive Eqs.~\eqref{eq:mean-efficiency} and \eqref{eq:variance-efficiency}, where the means and variances of the detected photon numbers are expressed by the same properties of the emitted photon numbers, respectively.  The means are straightforward to express with the expectation values of the number of emitted photons using Eq.~\eqref{eq:law-of-tot-exp}:
\begin{equation}
\begin{split}
    \mu_{0/1} = \sum_{i=0}^\infty P_{0/1}^\mathrm{e}(i)\mathrm{E}\left(n_{0/1}|n_{0/1}^\mathrm{e}=i\right)
    = \sum_{i=0}^\infty P_{0/1}^\mathrm{e}(i)i\eta = \mu_{0/1}^\mathrm{e}\eta.
    \label{eq:mu-eff}
\end{split}
\end{equation}
We used that if $i$ photons are emitted, then the detected photon number is a binomial random variable with $i$ trials and success probability $\eta$. The variances can be expressed similarly using Eq.~\eqref{eq:law-of-tot-var}:
\begin{widetext}
\begin{equation}
\begin{split}
    \sigma_{0/1}^2 &= \sum_{i=0}^\infty P_{0/1}^\mathrm{e}(i)\mathrm{Var}(n_{0/1}|n_{0/1}^\mathrm{e}=i) + \sum_{i=0}^\infty P_{0/1}^\mathrm{e}(i) \left(\mathrm{E}(n_{0/1}|n_{0/1}^\mathrm{e}=i)- \sum_{j=0}^\infty P_{0/1}^\mathrm{e}(j)\mathrm{E}(n_{0/1}|n_{0/1}^\mathrm{e}=j)\right)^2\\
    &= \eta(1-\eta)\mu_{0/1}^\mathrm{e} + \eta^2 (\sigma_{0/1}^\mathrm{e})^2.
    \label{eq:var-eff}
\end{split}
\end{equation}
\end{widetext}
The final expressions of Eqs.~\eqref{eq:mu-eff} and \eqref{eq:var-eff} are stated in Eqs.~\eqref{eq:mean-efficiency} and \eqref{eq:variance-efficiency} respectively.

\subsection{Calculating how the detected photon statistics depend on the initialization error}
The expectation values and variances of the detected photon distributions in the presence of initialization error can also be calculated using the law of total expectation and law of total variance respectively. In Eqs.~\eqref{eq:law-of-tot-exp} and \eqref{eq:law-of-tot-var}, the detected photon number $n_{0/1}$ takes the role of $X$, and $Y$ encodes whether or not an initialization error happened. A possible encoding is that $Y$ takes the value $0$ if no initialization error happened, which has probability $1-q$, and it takes the value $1$ if an initialization error happened, which has probability $q$. Using this notation, the expectation values can be expressed as
\begin{equation}
    \mu_{0/1} = \sum_{y\in\{0, 1\}} P(y)\mathrm{E}\left(n_{0/1}|Y=y\right) = (1-q)\mu_{\ket{0/1}}+q\mu_{\ket{1/0}}.
\end{equation}
The final expression is stated in Eq.~\eqref{eq:muinit}.
Similarly, the variance $\sigma_0^2$ is
\begin{widetext}
\begin{equation}
\begin{split}
    \sigma_{0}^2 &= \sum_{y\in\{0, 1\}}P(y)\mathrm{Var} \left(n_{\ket{0\oplus y}}\right) +\sum_{y\in\{0, 1\}} P(y) \left(\mathrm{E}(n_{\ket{0\oplus y}}) - \sum_{z\in\{0, 1\}} \mathrm{E}\left(n_{\ket{0\oplus z}}\right)P(z) \right)^2\\
    &= (1-q)\sigma_{\ket{0}}^2+ q\sigma_{\ket{1}}^2 + (1-q)\left\{\mu_{\ket{0}}-\left[(1-q)\mu_{\ket{0}}+q\mu_{\ket{1}}\right]\right\}^2 + q\left\{\mu_{\ket{1}}-\left[(1-q)\mu_{\ket{0}}+q\mu_{\ket{1}}\right]\right\}^2.
\end{split}
\end{equation}
\end{widetext}
Here, $\oplus$ denotes addition modulo 2.
The variance $\sigma_1^2$ is obtained by changing all $0$ indices to $1$ and vice versa. Expanding the parenthesis in the third expression leads to Eq.~\eqref{eq:sigmainit}.

\section{Photon-number resolved rate equation}
\label{app:rateequation}

The time dependence of the populations and the photon-counting statistics can be calculated using the photon-number-resolved rate equation; this is, in fact, an infinite set of coupled rate equations, as seen below.
Our treatment is based on Ref.~\onlinecite{Panadero}.

The level diagram, the transitions between the levels, as well as the rates, are shown in Fig.~\ref{fig:setup}(b).
In the rate equation, the unknown is the infinite vector $\boldsymbol{p}(t)$, whose entry $p_{\sigma,n}(t)$ is the probability of occupying level $\sigma$ at time $t$ after the switch-on of the laser pulse, such that $n$ photoluminescence photons have been emitted since the switch-on.
That is, the normalization condition reads
$\sum_{\sigma \in S} \sum_{n=0}^\infty p_{\sigma,n}(t)= 1$ for all times $t$.

The photon-number-resolved rate equation is the following infinite set of ordinary first-order differential equations:
\begin{widetext}
\begin{eqnarray}
    \dot{\boldsymbol{p}}_n &=& \boldsymbol{M}_{+} \boldsymbol{p}_{n-1} + \boldsymbol{M} \boldsymbol{p}_n,\\
    \boldsymbol{p}_n &=& \left(p_{\ket{\text{g}, -1}, n}; p_{\ket{\text{g}, 0}, n}; p_{\ket{\text{g}, 1}, n}; p_{\ket{\text{e}, -1}, n}; p_{\ket{\text{e}, 0}, n}; p_{\ket{\text{e}, 1}, n}; p_{\ket{\text{s}}, n} \right)^T,\\
    \boldsymbol{M}_{+}  &=&\begin{bmatrix}0 & 0 & 0 & \Gamma_{0} & 0 & 0 & 0\\0 & 0 & 0 & 0 & \Gamma_{0} & 0 & 0\\0 & 0 & 0 & 0 & 0 & \Gamma_{0} & 0\\0 & 0 & 0 & 0 & 0 & 0 & 0\\0 & 0 & 0 & 0 & 0 & 0 & 0\\0 & 0 & 0 & 0 & 0 & 0 & 0\\0 & 0 & 0 & 0 & 0 & 0 & 0\end{bmatrix},\\
    \boldsymbol{M} &=&\begin{bmatrix}
        - \Gamma_{P} & 0 & 0 & 0 & 0 & 0 & \Gamma_{s1}\\0 & - \Gamma_{P} & 0 & 0 & 0 & 0 & \Gamma_{s0}\\0 & 0 & - \Gamma_{P} & 0 & 0 & 0 & \Gamma_{s1}\\\Gamma_{P} & 0 & 0 & - \Gamma_{0} - \Gamma_{f1} & 0 & 0 & 0\\0 & \Gamma_{P} & 0 & 0 & - \Gamma_{0} - \Gamma_{f0} & 0 & 0\\0 & 0 & \Gamma_{P} & 0 & 0 & - \Gamma_{0} - \Gamma_{f1} & 0\\0 & 0 & 0 & \Gamma_{f1} & \Gamma_{f0} & \Gamma_{f1} & - \Gamma_{s0} - 2 \Gamma_{s1}\end{bmatrix}.
        \label{eq:master-eq}
\end{eqnarray}
\end{widetext}

\section{Measurement time dependence of the multi-shot readout error benchmark}
\label{app:meas-time-dependence}
\begin{figure}
    \centering
    \includegraphics[width=\linewidth]{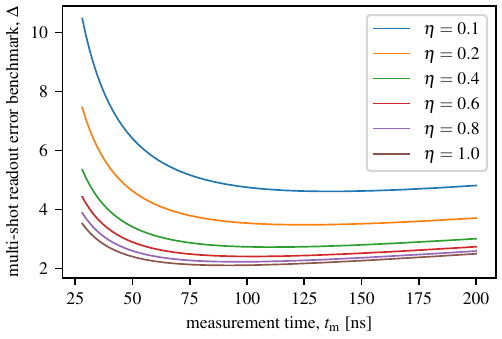}
    \caption{Multi-shot readout error benchmark as a function of measurement time. Parameters: see caption of Fig.~\ref{fig:setup} for rate values. No background photons ($\lambda=0$); and no initialization error ($q=0$).}
    \label{fig:delta-t}
\end{figure}

In Sec.~\ref{sec:nv}, we show how the multi-shot readout error benchmark depends on the photon detection efficiency, the background photon intensity and initialization error. The multi-shot readout error benchmark also depends on the measurement time. In Figs.~\ref{fig:MSE}(a) and \ref{fig:delta-q}, the multi-shot readout error benchmark is plotted at the optimal measurement time $\tau_\mathrm{m, opt}$ for which $\Delta$ is smallest. Here, we discuss the dependence of $\Delta$ on the measurement time, which is determined by solving the photon number resolved master equation in Eq.~\eqref{eq:master-eq}. This dependence is shown by Fig.~\ref{fig:delta-t}. The multi-shot measurement benchmark has a minimum at the optimal measurement time $t_\mathrm{m, opt}$, which depends on the photon collection efficiency $\eta$. Fig.~\ref{fig:MSE}(a) shows $\Delta$ at $t_\mathrm{m, opt}$, which is plotted in Fig.~\ref{fig:MSE}(b). Note that $t_\mathrm{m, opt}$ also depends on the background photon intensity $\lambda$, which is also shown by Fig.~\ref{fig:MSE}(b). Fig.~\ref{fig:delta-t} also shows that the dependence of $\Delta$ on $t_\mathrm{m}$ is small between 75~ns and 200 ns, moving away from $t_\mathrm{m, opt}$ in this range only changes $\Delta$ by a small amount.

%

\end{document}